# Tuning $J_{eff}$ = 1/2 Insulating State via Electron Doping and Pressure in Double-Layered Iridate $Sr_3Ir_2O_7$

L. Li[1], P. P. Kong[2], T. F. Qi[1], C. Q. Jin[2], S. J. Yuan[1,3], L. E. DeLong[1], P. Schlottmann[4] and G. Cao[1*]

[1]Department of Physics and Astronomy and Center for Advanced Materials, University of Kentucky, Lexington, KY 40506, USA

[2]Institute of Physics, Chinese Academy of Sciences, Beijing, China

[3]Department of Physics, Shanghai University, Shanghai, China

[4]Department of Physics, Florida State University, Tallahassee, FL 32306, USA

$Sr_3Ir_2O_7$ exhibits a novel $J_{eff}$=1/2 insulating state that features a splitting between $J_{eff}$=1/2 and 3/2 bands due to spin-orbit interaction. We report a metal-insulator transition in $Sr_3Ir_2O_7$ via either dilute electron doping ($La^{3+}$ for $Sr^{2+}$) or application of high pressure up to 35 GPa. Our study of single-crystal $Sr_3Ir_2O_7$ and $(Sr_{1-x}La_x)_3Ir_2O_7$ reveals that application of high hydrostatic pressure P leads to a drastic reduction in the electrical resistivity by as much as six orders of magnitude at a critical pressure, $P_C$ = 13.2 GPa, manifesting a closing of the gap; but further increasing P up to 35 GPa produces no fully metallic state at low temperatures, possibly as a consequence of localization due to a narrow distribution of bonding angles θ. In contrast, slight doping of $La^{3+}$ ions for $Sr^{2+}$ ions in $Sr_3Ir_2O_7$ readily induces a robust metallic state in the resistivity at low temperatures; the magnetic ordering temperature is significantly suppressed but remains finite for $(Sr_{0.95}La_{0.05})_3Ir_2O_7$ where the metallic state occurs. The results are discussed along with comparisons drawn with $Sr_2IrO_4$, a prototype of the $J_{eff}$ = 1/2 insulator.

I. Introduction

Traditional arguments suggest that iridium oxides should be more metallic and less magnetic than materials based upon 3*d* and 4*f* elements, because 5d-electron orbitals are more extended in space, which increases their electronic bandwidth. This conventional wisdom conflicts with two trends observed among layered iridates such as the Ruddlesden-Popper phases, $Sr_{n+1}Ir_nO_{3n+1}$ (n = 1 and 2; n defines the number of Ir-O layers in a unit cell) **[1-5]** and hexagonal $BaIrO_3$ **[6]**. First, complex magnetic states occur with high critical temperatures but unusually low ordered moments. Second, "exotic" insulating states are observed rather than metallic states **[4-6]** (see **Table 1**).

**Table 1.** *Examples of Layered Iridates*

| System | Néel/Curie Temperature (K) | Ground State |
|---|---|---|
| $Sr_2IrO_4$ (n =1) | 240 | *Canted AFM insulator* |
| $Sr_3Ir_2O_7$ (n =2) | 285 | *AFM insulator* |
| $BaIrO_3$ | 183 | *Magnetic insulator* |

A critical underlying mechanism for these unanticipated states is a strong spin-orbit interaction (SOI) that vigorously competes with Coulomb interactions, non-cubic crystalline electric fields, and Hund's rule coupling. The net result of this competition is to stabilize ground states with exotic behavior. The most profound effect of the SOI on the iridates is the $J_{eff}$ = 1/2 insulating state **[7-9],** a new quantum state that exemplifies novel physics in the 5d-electron systems. The SOI is a relativistic effect proportional to $Z^4$ (Z is the atomic number), and has an approximate strength of 0.4 eV in the iridates (compared to around 20 meV in 3d materials), and splits the $t_{2g}$ bands into states with $J_{eff}$ = 1/2 and $J_{eff}$ = 3/2, the latter having lower energy **[7, 8]** (see **Table 2**). Since $Ir^{4+}$ ($5d^5$) ions provide five 5d-electrons to bonding states, four of them fill the lower $J_{eff}$ = 3/2 bands, and one electron partially fills the $J_{eff}$ = 1/2 band where the Fermi level $E_F$ resides. The $J_{eff}$ = 1/2 band is so narrow that even a reduced on-site Coulomb repulsion (U ~

0.5 eV, due to the extended nature of 5d-electron orbitals) is sufficient to open a small gap $\Delta$ that stabilizes the insulating state in the iridates **[7, 8, 11, 12]**.

The splitting between the $J_{eff} = 1/2$ and $J_{eff} = 3/2$ bands narrows as the dimensionality (i.e., n) increases in $Sr_{n+1}Ir_nO_{3n+1}$, and the two bands progressively broaden and contribute to the density of states (DOS) near the Fermi surface. In particular, the bandwidth W of $J_{eff} = 1/2$ band increases from 0.48 eV for n = 1 to 0.56 eV for n = 2 and 1.01 eV for n = ∞ **[8, 11]**. The ground state evolves with decreasing $\Delta$, from a robust insulating state for $Sr_2IrO_4$ (n = 1) to a metallic state for $SrIrO_3$ (n = ∞) as n increases. A well-defined, yet weak, insulating state lies between them at $Sr_3Ir_2O_7$ (n = 2). Given the delicate balance between relevant interactions, a recent theoretical proposal predicts $Sr_3Ir_2O_7$ to be at the border between a collinear AFM insulator and a spin-orbit Mott insulator **[13]**.

**Table 2.** *Comparison between 3d and 4d/5d Electrons*

| Electron Type | U(eV) | $\lambda_{os}$(eV) | Spin State | Interactions | Phenomena |
|---|---|---|---|---|---|
| 3d | 5-7 | 0.01-0.1 | High | $U > CF > \lambda_{so}$ | Magnetism/HTSC |
| 4d | 0.5-3 | 0.1-0.3 | Intermediate | $U \sim CF > \lambda_{so}$ | Magnetism |
| 5d | 0.4-2 | 0.1-1 | Low | $U \sim CF \sim \lambda_{so}$ | $J_{eff} = 1/2$ State |

The onset of weak ferromagnetic (FM) order is observed at $T_C$ = 240 K **[4-7]** and 285 K **[8]** in the case of $Sr_2IrO_4$ and $Sr_3Ir_2O_7$, respectively (see **Fig. 1**). It is generally recognized that the magnetic ground state for $Sr_2IrO_4$ and $Sr_3Ir_2O_7$ (i.e., for both n = 1, 2) is antiferromagnetic (AFM) and is closely associated with the rotation of the $IrO_6$ octahedra about the **c**-axis, which characterizes the crystal structure of both $Sr_2IrO_4$ and $Sr_3Ir_2O_7$ **[1-3, 5]**. Indeed, the temperature dependence of the magnetization M(T) closely tracks the rotation of the octahedra, as characterized by the Ir-O-Ir bond angle θ, for both n = 1 and 2, as shown in **Figs. 1a** and **1b**. A recent neutron study of single-crystal $Sr_2IrO_4$ reveals a canted AFM structure in the basal plane with spins primarily aligned along the **a**-axis **[14]**. Unlike $Sr_2IrO_4$, $Sr_3Ir_2O_7$ exhibits an intriguing magnetization reversal in the **a**-axis magnetization $M_a(T)$ below $T_D$ = 50 K; both $T_C$ and $T_D$ can be observed only when the system is field-cooled (FC) from above $T_C$ (**Fig. 1b**) **[5]**. This behavior is robust and not observed in the zero-field cooled (ZFC) magnetization, which instead remains positive and displays no anomalies that are seen in the FC magnetization **[5]**. It is also noted that the **c**-axis magnetization $M_c(T)$ is much weaker and the magnetic ordering is not so well-defined (see **Fig. 1b**).

In spite of the extremely weak scale of the anomalies in M(T,H), corresponding anomalies in the electrical resistivity ρ and specific heat C(T) are observed at $T_C$ = 285 K in $Sr_3Ir_2O_7$. In sharp contrast, correlated anomalies in M(T), C(T) and ρ(T) at $T_C$ are either weak, or conspicuously absent in $Sr_2IrO_4$ **[4, 15]**, as illustrated in **Fig. 1**. For example, the observed specific heat anomaly |ΔC| ~ 10 J/mole K at $T_C$ for $Sr_3Ir_2O_7$ corresponds to a sizable entropy change; this change is tiny (~ 4 mJ/mole K) for $Sr_2IrO_4$, in spite of its robust, long-range magnetic order at $T_C$ = 240 K **[1-4, 10, 15-18]**. The weak phase transition signatures suggest that thermal and transport properties may not be driven by the same interactions that dictate the magnetic behavior in $Sr_2IrO_4$. Recently, a time-resolved optical study of indicates that $Sr_2IrO_4$ is a unique system in which Slater- and Mott-Hubbard-type behaviors coexist, which might explain the absence of anomalies at $T_C$ in transport and thermodynamic measurements **[19]**.

The nature of magnetism and its implications in the iridates are currently open to debate, in part because the spin degree of freedom is no longer an independent parameter, owing to the strong SOI in the iridates. More important, it is clear that the properties of these materials are strongly influenced by the lattice degrees of freedom **[15-18, 20]**.

The experimental record reviewed above proves that $Sr_3Ir_2O_7$ provides a unique prototype for the novel ground states open to iridates. In this paper, we report results of our structural, transport, magnetic and thermal study of single-crystal $Sr_3Ir_2O_7$ as a function of pressure up to 35 GPa and slight La doping for Sr. This study reveals that application of high hydrostatic pressure P leads to a drastic reduction in the electrical resistivity $\rho$ by up to six orders of magnitude; a sudden drop in $\rho$ at a critical pressure, $P_C = 13.2$ GPa, is observed. However, a nonmetallic state remains below 5 K even at P up to 35 GPa. In contrast, a robust metallic state throughout the entire measured temperature range is readily induced by mere 5% doping of $La^{3+}$ ions for $Sr^{2+}$ ions in $Sr_3Ir_2O_7$ or in $(Sr_{0.95}La_{0.05})_3Ir_2O_7$; furthermore, the a-axis resistivity $\rho_a$ has a rapid drop below 20 K. The metallic state is accompanied by a significant increase in the Ir-O-Ir bond angle $\theta$, particularly at low temperatures; this correlation highlights a critical role of $\theta$ in determining the ground state. Remarkably, electron doping considerably weakens the magnetic state without completely suppressing $T_C$ in a fully established metallic state in $(Sr_{0.95}La_{0.05})_3Ir_2O_7$. This behavior sharply contrasts that for La-doped $Sr_2IrO_4$, where magnetic order is completely suppressed in the metallic state **[17]**.

**II. Experimental**

Single crystals studied were synthesized using a self-flux technique described elsewhere **[7, 8, 15-17]**. The average size of the single crystals is 0.5 x 0.5 x 0.1 $mm^3$. The structures of $(Sr_{1-x}La_x)_3Ir_2O_7$ were determined using a Nonius Kappa CCD X-Ray Diffractometer with sample temperature controlled using a nitrogen stream; they were refined by full-matrix least-squares method using the SHELX-97 programs **[21]**. Chemical compositions of the single crystals were determined using energy dispersive X-ray analysis (EDX) (Hitachi/Oxford 3000). Resistivity, magnetization and specific heat were measured using a Quantum Design MPMS7 SQUID Magnetometer and a Quantum Design Physical Property Measurement System with 14 T field capability. High-pressure measurements of $\rho(T)$ were carried out using a CuBe diamond anvil cell (DAC) with thin Au wires as electrodes and a fine powder of hexagonal BN as the pressure medium, as described in **Ref. 22**.

The crystal structure of $Sr_3Ir_2O_7$ features a rotation of the $IrO_6$-octahedra about the c-axis, resulting in a larger unit cell by √2 x √2 x 2 [5]. Our single-crystal X-ray refinement confirms that this rotation corresponds to a distorted in-plane Ir-O-Ir bond angle θ, whose value is sensitive to temperature. Values of the representative bond angle θ for $(Sr_{1-x}La_x)_3Ir_2O_7$ are listed in **Table 3 (where Δθ = θ(295 K) - θ(90 K))**, and the definition of θ is given in an inset to **Fig. 1b**. It is clear that La doping reduces the lattice distortion by significantly increasing θ particularly at low temperatures.

**Table 3** *The bond angle Ir-O-Ir θ at 90 K and 295K*

| x | θ (deg) at 90K | θ (deg) at 295K | Δθ (deg) |
|---|---|---|---|
| 0 | 154.922 | 156.500 | 1.578 |
| 0.05 | 156.034 | 156.523 | 0.489 |

**III. Results and Discussion**

Hopping between active $t_{2g}$ orbitals is critically linked to the Ir-O-Ir bond angle θ, as manifested in the electrical resistivity $\rho(T)$ of single-crystal $(Sr_{1-x}La_x)_3Ir_2O_7$, as shown in **Fig. 2**. The **a**-axis resistivity $\rho_a$ (the **c**-axis resistivity $\rho_c$) is reduced by as much as a factor of $10^{-6}$ ($10^{-5}$) at low T as x is increased from 0 to 0.05 (see **Figs. 2a and 2b**). For x = 0.05, there is a sharp downturn in $\rho_a$ near 20 K, indicative of a rapid decrease in elastic scattering (**Fig. 2c**). Such low-T behavior is also observed in slightly oxygen depleted $Sr_2IrO_{4-\delta}$, with δ = 0.04 and La-doped $Sr_2IrO_4$ [17, 18]. The radical changes in the transport properties of $Sr_3Ir_2O_7$ with slight doping can arise from three effects: (1) there is chemical pressure, because the size of $La^{3+}$ is smaller than that of $Sr^{2+}$ ions, (2) doping, since the different valence of the ions adds electrons to system, and (3) disorder scattering. Disorder breaks the translational invariance of the lattice and tends to develop tails in the gap (it becomes a pseudo-gap). These states are likely to be localized at low T, i.e. they do not contribute to conduction. The additional electrons due to doping fill the states in the gap and push the Fermi level into the bottom of the upper $J_{eff}$ = ½ Hubbard band. The conduction now depends on the relative position of the Fermi level with respect to the mobility edge. For small x the compound is expected to be insulating and become metallic for larger x. The effect of chemical pressure is to reduce the gap, because the hopping matrix element should increase and the availability of conduction electrons enhances the screening of the Coulomb interaction.

The inducement of such a robust metallic state by impurity doping further reinforces the central finding of this work: that is, transport properties in the iridates such as $Sr_3Ir_2O_7$ can be mainly dictated by the lattice degrees of freedom since electron hopping sensitively depends on the bond angle θ. Indeed, it is theoretically anticipated that hopping occurs through two active $t_{2g}$ orbitals, the $d_{xy}$ and $d_{xz}$ when θ = 180°, and the $d_{xz}$ and $d_{yz}$ orbitals when θ = 90° [20]. The data shown in **Fig. 2** demonstrate that the larger θ, the more energetically favorable it is for electron hopping and superexchange interactions (see **Table 3**).

It is noteworthy that the magnetically ordered state weakens considerably and $T_C$ decreases, with La doping in $(Sr_{1-x}La_x)_3Ir_2O_7$, but does not vanish at x = 0.05, where the metallic state is fully established. The transition at $T_C$ is broadened by doping in the **a**-axis magnetic susceptibility data $\chi_a(T)$, but is hardly visible in the **c**-axis magnetic susceptibility data $\chi_c(T)$, as shown in **Fig. 3**. The magnetic susceptibility appears to show correlation with the resistivity; for example, $\chi_a$(T) for x=0.05 shows a slope change below 20 K and near 220 K, respectively; and both $d\rho_a/dT$ and $d\rho_c/dT$ exhibit a corresponding slope change below 20 K, and $d\rho_c/dT$ also shows an anomaly near 220 K, as shown in **Fig. 2c**. This behavior sharply contrasts that in La-doped $Sr_2IrO_4$, where the occurrence of a fully metallic state is accompanied by the disappearance of magnetic order **[17]**. This difference could be attributed, in part, to the fact that $Sr_3Ir_2O_7$ with U ~ 0.5 eV is much closer to the borderline of the metal-insulator transition (MIT). Due to the proximity to the MIT, inhomogeneities in the sample are quite likely, which manifest themselves in a narrow distribution for the angle θ, and hence to inhomogeneities in the magnetization.

On the other hand, $Sr_2IrO_4$ has an energy gap as large as 620 meV **[12]** but a relatively small magnetic coupling energy of 60-100 meV **[23, 24]** that cannot significantly affect the insulating state. This may in part explain the lack of the correlation between transport and magnetic properties that characterizes $Sr_2IrO_4$. The differences between the two systems indicate that magnetic ordering plays different roles in determining the electronic state of either $Sr_3Ir_2O_7$ or $Sr_2IrO_4$. Indeed, a recent neutron diffraction study on single-crystal $Sr_2IrO_4$ reveals a canted spin structure within the basal plane at temperatures below $T_C$, and forbidden nuclear reflections of space group *I41/acd* appear over a wide temperature range from 4 K to 600 K, which indicates a reduced crystal symmetry **[14]**. In contrast, the magnetic structure of $Sr_3Ir_2O_7$ is believed to be a collinear AFM state along the **c**-axis; but there may exist a nearly degenerate magnetic state with canted spins in the basal plane **[13, 16, 25, 26]**.

The high sensitivity of the ground state to the lattice degrees of freedom strongly suggests that application of pressure will provides an effective probe of the insulating state of the iridates. Indeed, $\rho_a$(T) undergoes a drastic reduction by six orders of magnitude at a critical pressure, $P_C =$ 13.2 GPa, which marks a transition from an insulating state to a nearly metallic state in $Sr_3Ir_2O_7$, as shown in **Fig. 4a**. Furthermore, a broad transition between 210 K and 250 K in $\rho_a(T)$ is observed at $P_C$. It is noted that the nonmetallic behavior remains below 10 K, although the magnitude of $\rho_a$ is drastically reduced; and further increases of P up to 35 GPa do not significantly improve the metallic behavior. For example, $\rho_a$ at P = 25 GPa (>$P_C$) features a very weak temperature dependence at high temperatures that is then followed by an abrupt upturn or a transition in $\rho_a$ at 5 K, as shown in the inset in **Fig. 4a**. $\rho_a(T)$ approximately follows an activation law, $\rho_a (T) \sim \exp(\Delta/2k_BT)$ (where $\Delta$ is the energy gap and $k_B$ the Boltzmann's constant) in the temperature interval 10 < T < 100 K that reflects very small values of $\Delta$; the pressure dependence of $\Delta$ confirms that $P_C$ (= 13.2 GPa) indeed marks a transition from an insulating state to a nearly metallic state. This transition is also corroborated by the fact that the ratio $\rho_a(2K)/\rho_a(300K)$, which reflects the localization of electrons, closely tracks $\Delta$, as shown in **Fig. 4b**. A crude estimate suggests that 10 GPa is approximately equivalent to 100 meV/Å$^3$ **[27]**, therefore the lack of a fully metallic state under the high pressure in this borderline insulator is remarkably unusual.

A possible explanation for the remaining low-T insulating behavior is pressure-induced inhomogeneities (pressure gradients) leading to a narrow distribution of values for the angle θ. This breaks the translational invariance leading to localization over a small range of energy about the Fermi level. This only affects the low T resistivity, since for temperatures larger than the localization energy range the system behaves like a metal.

Interestingly, a plot of $\rho_a(2K)$ as a function of x for $(Sr_{1-x}La_x)_3Ir_2O_7$ and pressure P for $Sr_3Ir_2O_7$ seems to suggest that the effect of pressure near $P_C$ = 13.2 GPa is roughly equivalent to that of La doping near x = 0.03, as shown in **Fig. 5**. For $Sr_2IrO_4$, application of pressure also significantly reduces the electrical resistance by a few orders of magnitude near 20 GPa (where the weak ferromagnetism disappears), but the ground state remains insulating up to 40 GPa **[10]**, owing to a significantly wider insulating gap in the quasi-two-dimensional system (~ 620 meV) **[7, 8, 11, 12]**.

In addition, it deserves to be pointed out that the observed anisotropic diamagnetism (see **Fig. 1b**) can be attributed to the large SOI that breaks spin conservation. A wave-function $\Psi_\mathbf{k}$ of wave-vector **k** and spin state $|\uparrow\rangle$ or $|\downarrow\rangle$ is then of the form $\Psi_\mathbf{k} = a_\mathbf{k} |\uparrow\rangle + b_\mathbf{k} |\downarrow\rangle$, where $|a_\mathbf{k}|^2 + |b_\mathbf{k}|^2 = 1$, and the magnetization $M = \frac{1}{2} g\mu_B H \Sigma_\mathbf{k}(|a_\mathbf{k}|^2 - |b_\mathbf{k}|^2)$. Note the net moment M can be strongly reduced with respect to the usual Pauli susceptibility, and can even be negative. If the standard Landau susceptibility (quantization of orbits) and the core diamagnetism are added to the spin susceptibility, it is then likely to result in overall diamagnetism. Since this effect is only observed for FC but not for ZFC, we must conclude that the angle θ depends on the applied magnetic field and has a memory of the recent field history (similar to a glass).

## IV. Conclusions

The spin-orbit interaction vigorously competes with Coulomb interactions, non-cubic crystal electric field interactions, and the Hund's rule coupling in the iridates. In particular, traditional arguments would predict that iridates should have a metallic ground state and display Fermi liquid properties. However, the metallic state is rarely observed in the iridates, and does not exhibit Fermi-liquid behavior at low temperatures when realized. Indeed, the insulating state is often attained in the iridates. $Sr_3Ir_2O_7$ is a magnetic insulator that is close to the MIT. Consequently, the ground state is highly susceptible to small external perturbations such as chemical doping, high pressures and magnetic field.

A number of unexpected properties of iridates have been explained in terms of a recently predicted, $J_{eff}= 1/2$ insulating state that is a profound manifestation of the SOI. Because of the strong SOI, spin is no longer a good quantum number, which is manifest in the unusual magnetic properties of $Sr_2IrO_4$ and $Sr_3Ir_2O_7$. It is particularly noteworthy that the magnetic properties of the iridates do not appear to be as closely associated with electric transport properties as it is the case in 3d transition metal oxides. Chemical doping is the most effective way to achieve a metallic state in iridates via increasing the carrier concentration, i.e. shifting the Fermi level, and increasing the Ir-O-Ir bond angle in the $IrO_6$ octahedra, which points to the importance of electron-lattice couplings and ultimately, the strong SOI. It is surprising that, in contrast to magnetic materials based on 3d elements, high pressure is not very effective in inducing a metallic state in iridates.

## Acknowledgments

This work was support by the U. S. National Science Foundation under grants DMR-0856234 and EPS-0814194, and the U. S. Department of Energy Office of Science, Basic Energy Sciences Grant No. DE-FG02-97ER45653. PS is supported by the U.S. Dept. of Energy under Grant No. DE-FG02-98ER45707.

*Corresponding author; email: cao@uky.edu*

**Captions**

**Fig.1.** Temperature dependence of: **(a)** the magnetization and Ir-O-Ir bond angle $\theta$ (right scale) for $Sr_2IrO_4$; **(b)** the magnetization M and Ir-O-Ir bond angle $\theta$ (right scale) for $Sr_3Ir_2O_7$; **(c)** the c-axis resistivity $\rho_c$ for $Sr_2IrO_4$ and $Sr_3Ir_2O_7$; and **(d)** the specific heat C for $Sr_2IrO_4$ and $Sr_3Ir_2O_7$.

**Fig.2.** Temperature dependence of **(a)** the **a**-axis resistivity $\rho_a$; and **(b)** the **c**-axis resistivity $\rho_c$ for $(Sr_{1-x}La_x)_3Ir_2O_7$. **(c)** Temperature dependence of $\rho_a$, $\rho_c$ and $d\rho_a/dT$ and $d\rho_c/dT$ (right scale) for x = 0.05

**Fig.3.** Temperature dependence of (a) the **a**-axis magnetic susceptibilty $\chi_a$; and (b) the **c**-axis magnetic susceptibility $\chi_c$ for $(Sr_{1-x}La_x)_3Ir_2O_7$. Note that no magnetic anomalies are observed in $\chi_c$

**Fig.4**. **(a)** Temperature dependence of the **a**-axis resistivity $\rho_a$ at representative pressures P for $Sr_3Ir_2O_7$. Inset: $\rho_a$ vs. T for 25.5 GPa. **(b)** The activation gap $\Delta$ (left scale) and ratio $\rho_a$ (2 K)/$\rho_a$ (300K) (right scale) versus P for $Sr_3Ir_2O_7$. Inset: Schematic evolution of the density of states under pressure.

**Fig. 5**. The **a**-axis resistivity $\rho_a$ at 2 K as a function of x for $(Sr_{1-x}La_x)_3Ir_2O_7$ and pressure (upper axis) for $Sr_3Ir_2O_7$.

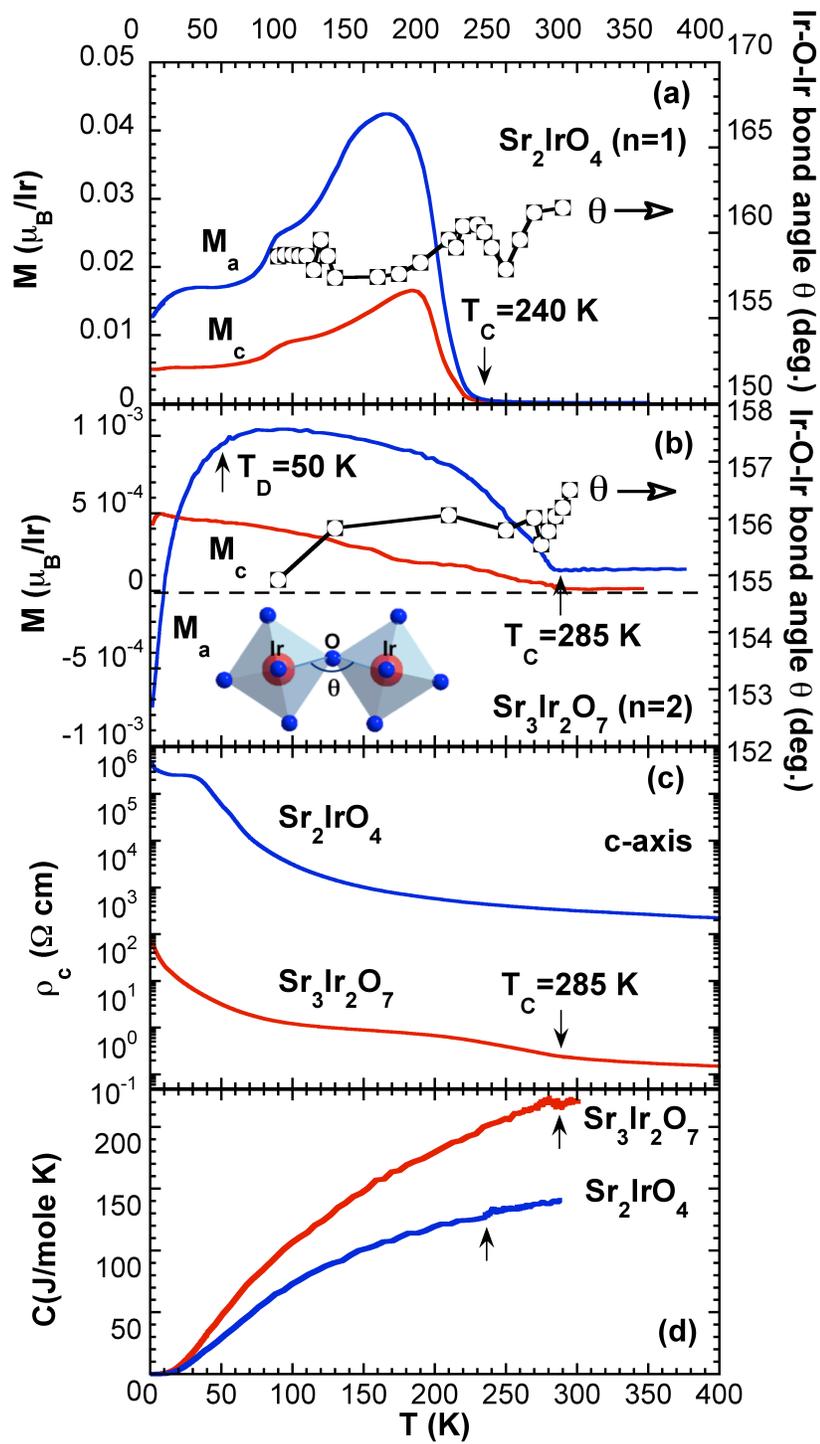

Fig. 1

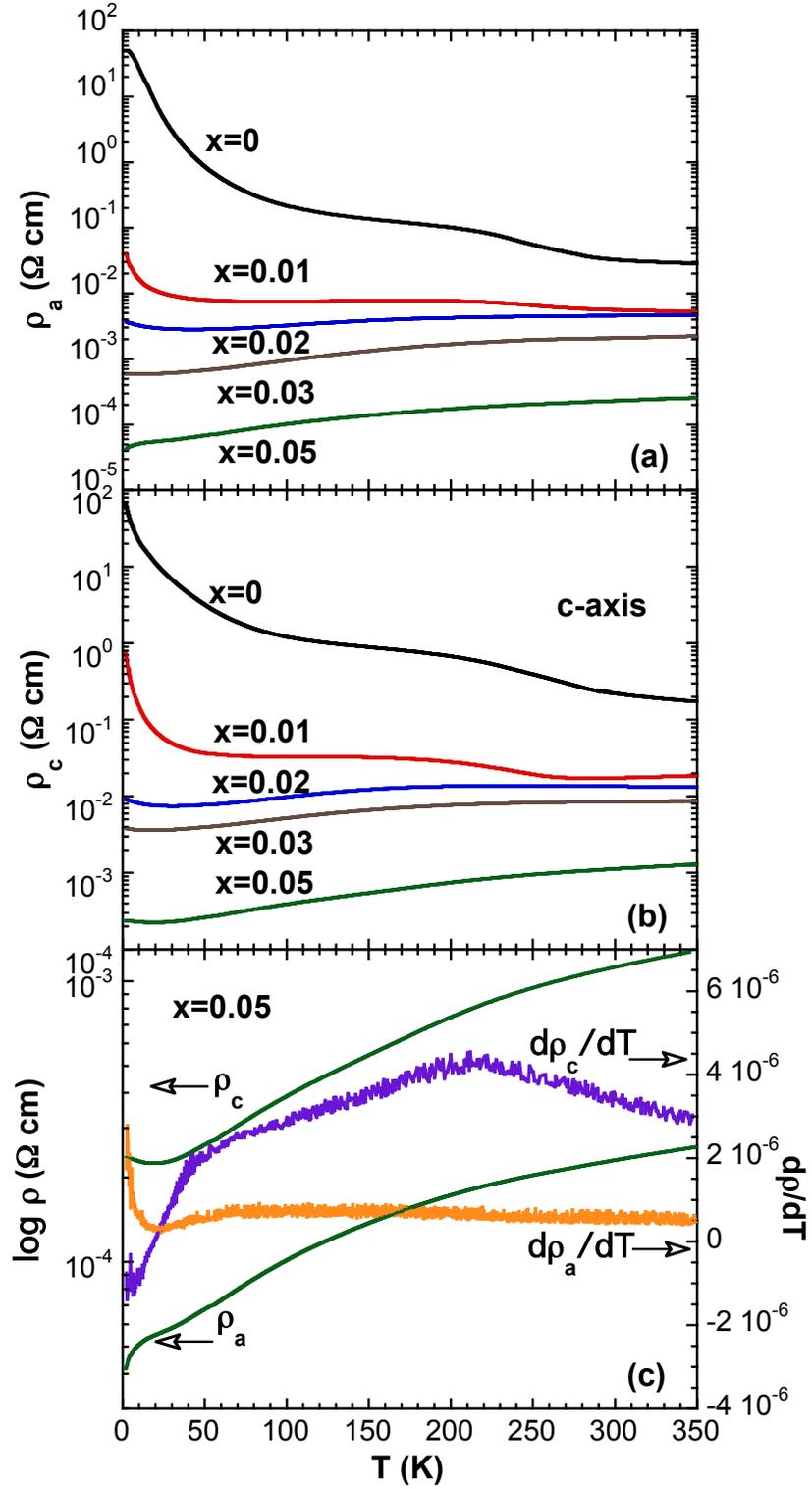

Fig. 2

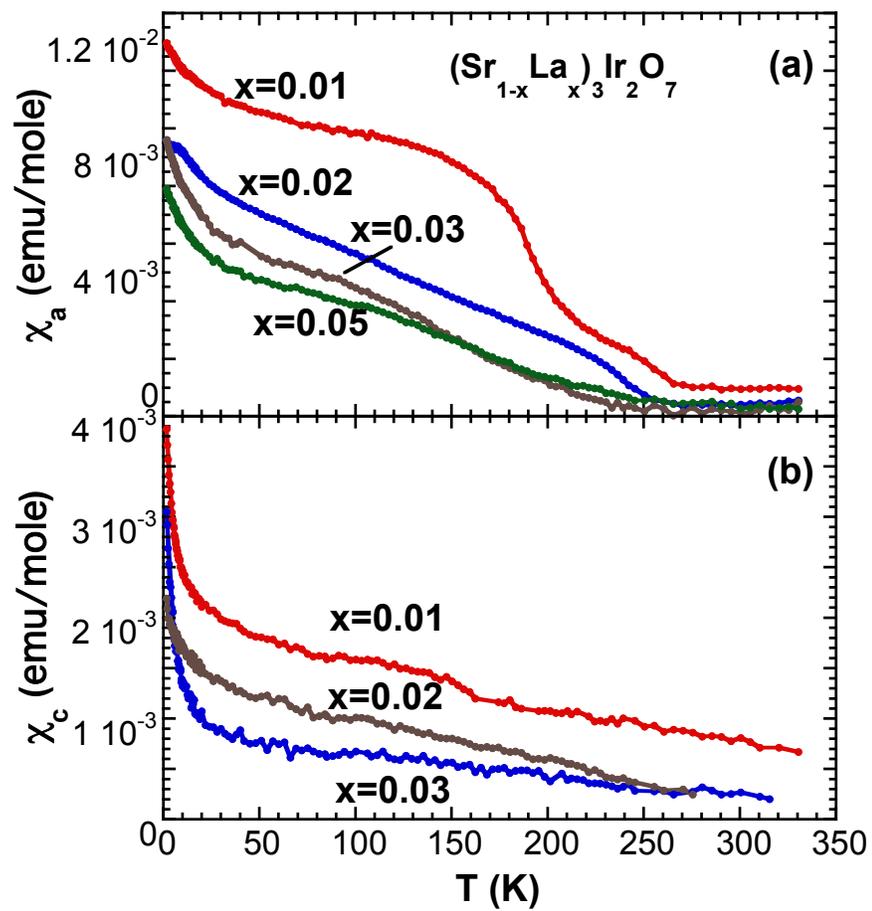

Fig. 3

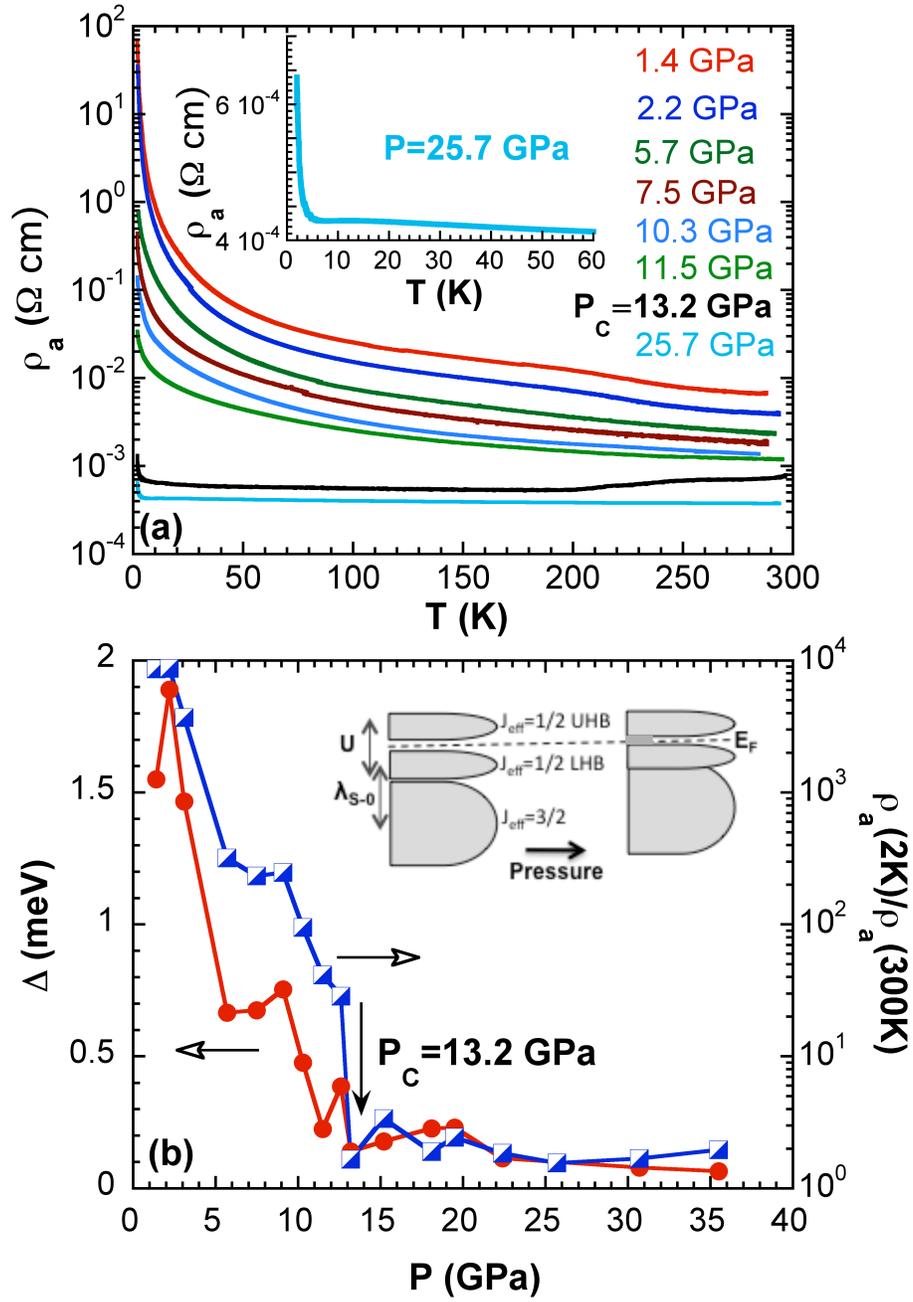

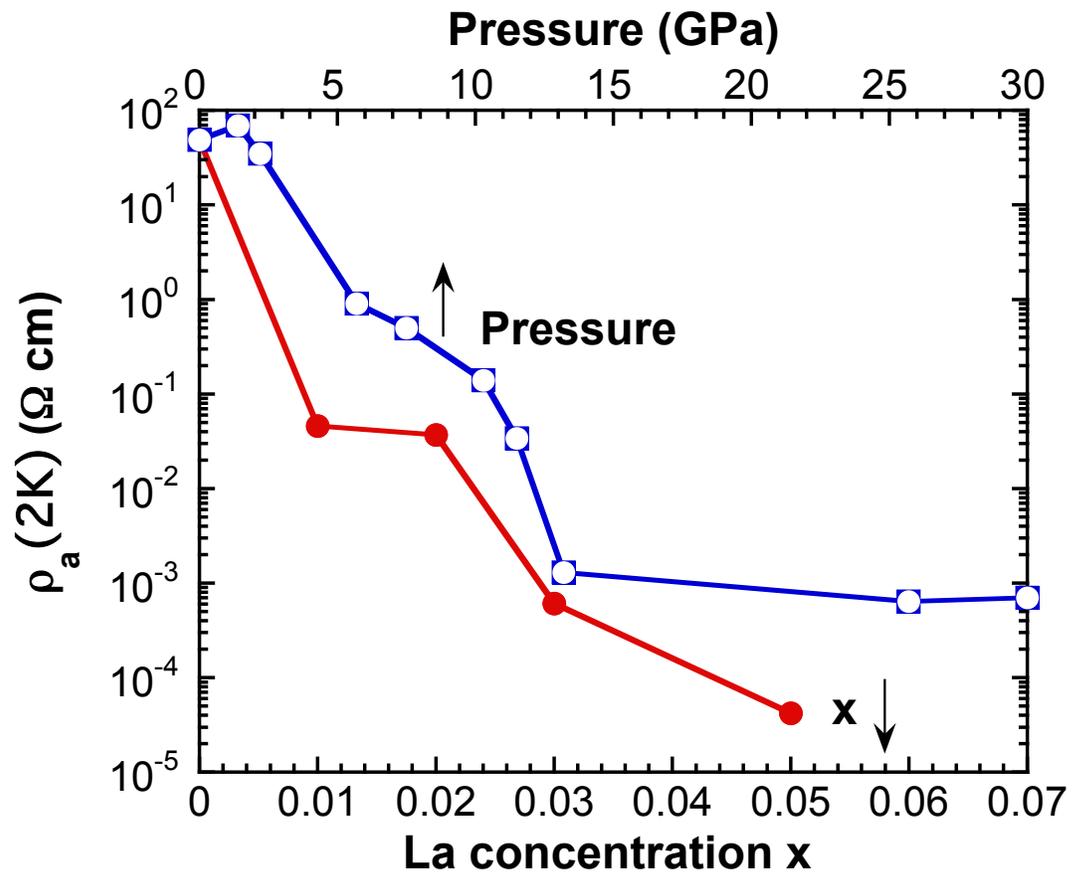

Fig. 5